\documentclass[aps]{revtex4}

\usepackage{amsmath,amsfonts,amssymb}
\usepackage{graphicx}
\usepackage[colorlinks=true, allcolors=blue]{hyperref}

\usepackage{ulem}
\usepackage{float}
\usepackage{color}

\begin{document} 

\title{Effects of resonances and surface texturing on light emission in emerging thin-film devices}

\author{Pyry Kivisaari$^a$}
\email[Corresponding author. Email: ]{pyry.kivisaari@aalto.fi}
\author{Mikko Partanen$^b$}
\author{Jani Oksanen$^a$}
\affiliation{$^a$Engineered Nanosystems Group, Aalto University, P.O. Box 13500, FI-00076 Aalto, Finland}
\affiliation{$^b$Department of Electronics and Nanoengineering, Aalto University, P.O. Box 13500, FI-00076 Aalto, Finland}

\begin{abstract}
Recent developments in thin-film fabrication and processing open up interesting possibilities for both established and emerging optics technologies. There, one of the key questions requiring more complete understanding is by how much one can improve the performance of thin-film devices by utilizing resonance effects and surface texturation. In this work, we report on our recent theoretical investigations around two aspects of this question: (1) how much the overall (=angle and energy-integrated) emission of extremely thin ($\sim10$ nm) layers can be enhanced through cavity effects, and (2) how much resonances affect the emission of moderately thin ($>100$ nm) layers in a typical device interacting with free space (in this case an ultra-thin solar cell). Beginning with topic (1), we find that the total emission of active layers with thicknesses $<50$ nm in particular can be boosted through resonant effects by placing them in a cavity. For topic (2), the results indicate that a radiative transfer approach (i.e., one not accounting for resonant effects) can give very accurate predictions of the total emission of moderately thin layers in a thin-film device, as long as the reflectances of the device's outer boundaries are known, and the emitting layer is not very close to optical elements supporting direct evanescent coupling (such as metal mirrors). Finally, we demonstrate that extending the self-consistent radiative transfer--drift-diffusion approach for diffusive scattering presents an interesting tool to optimize thin-film devices even with textured surfaces.
\end{abstract}

\maketitle

\section{INTRODUCTION}
\label{sec:intro}

Optical conversion and transmission of power in semiconductors is what solar photovoltaics is all about, but it is also at the heart of many emerging fields of optics, including photonic power transfer, optical DC converters, and thermophotonics \cite{zhao_2021,helmers_2021,sadi_2020,chen_2019}. All of these fields are also strongly influenced by recent developments in thin-film device fabrication and processing \cite{haggren_2023,schon_2022}. As the device layers are thus being shrunk towards the wavelength of light and even beyond, one of the key questions is by how much one can one boost the overall optical conversion efficiencies by making use of near-field and optical cavity effects \cite{chapuis_2023,song_2022,mittapally_2021,mcsherry_2019}. Even in more conventional solar photovoltaics, where ultra-thin solar cells are being explored as viable alternatives for the dominant silicon solar cells, it is of interest how much one can exploit resonant effects and surface texturing to optimize the performance of ultra-thin solar cells \cite{vaneerden_2019}.

Recently, we carried out fluctuational electrodynamics (FED) calculations and reported that optical cavities can substantially modify the total power emitted by a thin layer \cite{kivisaari_2022}. However, in that work we did not consider whether those modifications are sufficient for overcoming the total (=angle and energy-integrated) emission in the case where these thin layers emit into free space without any cavity effects. On the other hand, we recently introduced an interference-extended radiative transfer model to solve FED self-consistently with the position-dependent quasi-Fermi levels from electrical transport modeling \cite{partanen_2017_irt,kivisaari_2021}. The results revealed peculiar net photon recycling processes taking place in thin-film solar cells, but we did not study how much resonances affected the total magnitude of the phenomenon.

In this paper, we report on our ongoing investigations around the questions brought up in the previous paragraph. We focus on two general topics: (1) extremely thin ($\sim$ 10 nm) emitter and absorber layers enclosed in the same optical cavity, and (2) moderately thin ($>$100 nm) active layers as part of a device interacting with free space, in this case an ultra-thin solar cell. For topic (1), we study whether the overall electroluminescence (EL) of an extremely thin layer can be enhanced by placing it in an optical cavity together with an absorber layer. The simulations are carried out with and without a vacuum nanogap placed between the emitter and absorber parts of the cavity (vacuum gap is necessary for thermophotonic applications, where thermal insulation between the emitter and absorber is required). Also, the effect of the absorber layer thickness on the emission is studied. For topic (2), we study how much resonances affect the EL of moderately thin layers ($>$100 nm) in a solar cell by carrying out two comparative electro-optical simulations with the same parameters: one based on conventional radiative transfer (RT) and another one based on fluctuational electrodynamics (through the interference-extended RT model deployed also in Ref. \citenum{kivisaari_2021}).

The contents of this paper are as follows. Section \ref{sec:theory} summarizes the theoretical models deployed in the calculations of this paper, with more details on specific aspects given in Appendices \ref{sec:appendixa} and \ref{sec:appendixb}. Section \ref{sec:results} presents results on the two topics described above in dedicated Subsections: Section \ref{sec:cavities} presents results on how much the emission of extremely thin layers can be boosted with cavity effects, as compared to a reference case for emission into semi-infinite claddings; Section \ref{sec:etsolarcells} then compares with each other conventional and interference-extended radiative transfer simulations (coupled with the drift-diffusion model) for ultra-thin solar cells, first with specularly scattering surfaces. For greater technological relevance, results in Section \ref{sec:etsolarcells} are complemented with calculations carried out assuming a perfectly diffusively scattering back mirror. Appendices \ref{sec:appendixc} and \ref{sec:appendixd} give complementary results on top of the main results presented in Sections \ref{sec:cavities} and \ref{sec:etsolarcells}, respectively.

\section{THEORY} \label{sec:theory}

In all the calculations of this paper, the net radiative recombination-generation (RG) rate within the device is calculated as
\begin{equation}
R_{rad}(z) = \int_0^{\infty}d(\hbar\omega)\int_0^{\infty}dKr(z,K,\omega)K.
\label{eq:Rrad}
\end{equation}
Here, $z$ is position in the perpendicular direction with respect to the layer structure, $\hbar$ is the reduced Planck's constant, $\omega$ is the angular frequency (hence, $\hbar\omega$ is photon energy), $K$ is the magnitude of the in-plane component of the $\mathbf{k}$ vector, and $r$ is the spectral RG rate calculated from the spectral radiance $s$ (in Section \ref{sec:cavities}, $r=ds/dz$ so that $R_{rad}$ has units of W/m$^{3}$, and in Section \ref{sec:etsolarcells}, $r=\frac{1}{\hbar\omega}ds/dz$ so that $R_{rad}$ has units of 1/(m$^{3}$s) to comply with the drift-diffusion model to be presented below). The spectral radiance in turn is calculated from rightward and leftward propagating photon numbers $\phi^+$ and $\phi^-$ as
\begin{equation}
s(z,K,\omega) = \frac{1}{2}\sum_{TE,TM}\hbar\omega v(z,\omega)\mathcal{G}(z,K,\omega)[\phi^+(z,K,\omega)-\phi^-(z,K,\omega)].
\label{eq:xi}
\end{equation}
Here, $v(z,\omega)$ is the speed of light in the medium, and $\mathcal{G}(z,K,\omega)$ the direction-dependent optical density of states specified in Appendix \ref{sec:appendixa}.

In the calculations of optical cavities in Sec. \ref{sec:cavities}, the photon numbers are calculated by making direct use of the FED description deployed in Ref. \citenum{kivisaari_2022}. This description is well suited for calculating the net emission of a single layer assumed to have a given temperature and quasi-Fermi level separation. In Sec. \ref{sec:etsolarcells}, Eq. (\ref{eq:Rrad}) is instead solved iteratively with the drift-diffusion equations to arrive at self-consistent electro-optical solutions. To do this, it is beneficial to have an optical model which directly allows a device-wide and position-dependent excitation level that can even change inside single material layers. For this, we define a generalized radiative transfer (RT) equation for the photon numbers written as
\begin{align}
\frac{d}{dz}\phi^{\pm}(z,K,\omega) = & \mp\alpha(z,K,\omega)[\phi^{\pm}(z,K,\omega)-\eta(z,\omega)] \nonumber \\
	& \pm\beta(z,K,\omega)[\phi^{\mp}(z,K,\omega)-\eta(z,\omega)].
\label{eq:RT_gen}
\end{align}
Two versions of this generalized RT model are compared against each other in this work: (1) the conventional RT model \cite{chandrasekhar_1960} that does not include any interference effects in the actual calculations (for the outer boundaries, angle- and energy-dependent reflectances are calculated from an equivalent of the customary transfer matrix method \cite{kivisaari_2018}), and (2) the interference-extended radiative transfer (IRT) model of Ref. \citenum{kivisaari_2021} that provides a rigorous solution of FED. Coupled with the drift-diffusion (DD) model presented below, both RT and IRT models provide net recombination rates that also include the impact of photon recycling.

In Eq. (\ref{eq:RT_gen}), $\eta(z,\omega)$ is the Bose-Einstein distribution function given by
\begin{equation}
\eta(z,\omega)=\frac{1}{e^{[\hbar\omega-e\Delta E_F(z)]/[k_BT]}-1},
\end{equation}
where the quasi-Fermi level splitting $\Delta E_F(z) = E_{Fn}(z)-E_{Fp}(z)$ is used as the photon chemical potential in accordance with Ref. \citenum{wurfel} ($E_{Fn}$ and $E_{Fp}$ are defined in the next paragraph). Other variables required for solving Eqs. (\ref{eq:xi})--(\ref{eq:RT_gen}) are specified for the different models in Appendix \ref{sec:appendixa}.

For the sake of completeness, the DD model deployed in Section \ref{sec:etsolarcells} is given by
\begin{equation}
\begin{array}{c}
\frac{d}{dz}\left(-\varepsilon\frac{d}{dz} U\right)=e\left(p-n+\mathcal{N}\right),\\
\frac{d}{dz}J_{n}=\frac{d}{dz}\left(\mu_{n}n\frac{d}{dz} E_{Fn}\right)=e(R_{rad}-G_s+R_{nr}),\\
\frac{d}{dz}J_{p}=\frac{d}{dz}\left(\mu_{p}p\frac{d}{dz} E_{Fp}\right)=-e(R_{rad}-G_s+R_{nr}).
\end{array}\label{eq:DDyhtalot}
\end{equation}
Here, $\varepsilon$ is the static permittivity, $U$ is the electrostatic potential, $e$ is the elementary charge, $n$ and $p$ are electron and hole densities, $\mathcal{N}$ is the ionized doping density, $J_n$ and $J_p$ are the electron and hole current densities, $\mu_n$ and $\mu_p$ are the electron and hole mobilities, $E_{Fn}$ and $E_{Fp}$ are the conduction and valence band quasi-Fermi levels, $R_{rad}$ is the net radiative recombination-generation rate of the device calculated from Eq. (\ref{eq:Rrad}), $G_s$ is the generation rate due to incoming solar radiation calculated through separate photon number variables similarly as in Ref. \citenum{kivisaari_2021}, and $R_{nr}$ is the non-radiative recombination rate. The thus resulting models are called RTDD and IRTDD, corresponding to the conventional RT model and the more sophisticated IRT model, respectively.

The models specified so far have their standard solutions for perfectly planar geometries corresponding to specular reflection and transmission. However, the conventional RT model is straightforward to modify such that it accounts also for mirrors that scatter light diffusively, as described in Refs. \citenum{heikkila_2011_spie} and \citenum{heikkila_2011}. Selected RT model calculations are therefore carried out also for a perfectly scattering mirror in this work, with the corresponding equations presented in Appendix \ref{sec:appendixb}.

\section{RESULTS \& DISCUSSION} \label{sec:results}

\subsection{Emission boost in double-sided cavities} \label{sec:cavities}

\begin{figure}[htb]
\centering
\includegraphics[width=0.95\textwidth]{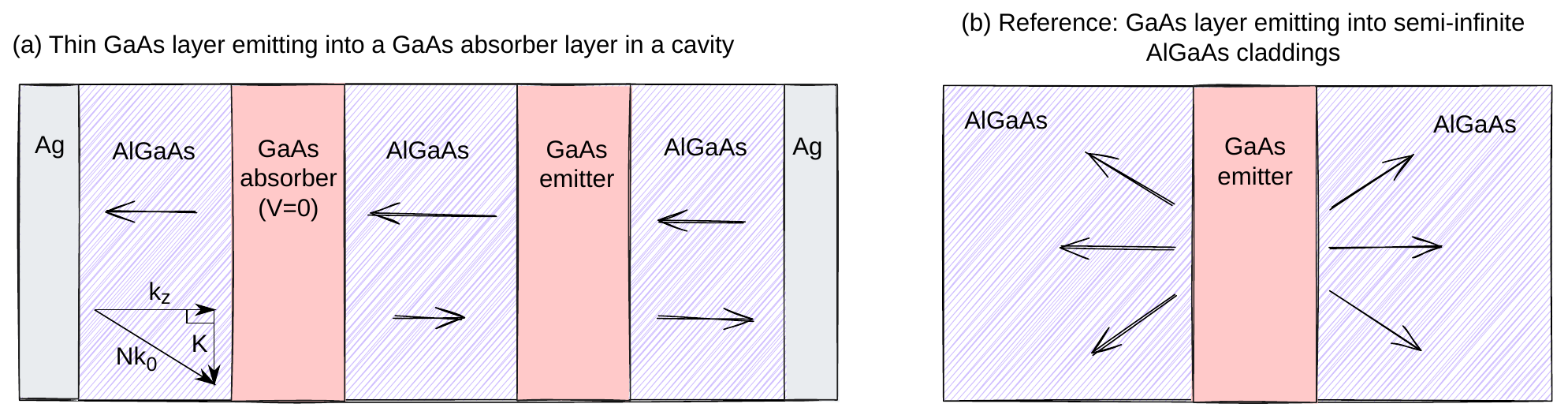}
\caption{(a) Intracavity structure consisting of GaAs emitter and absorber layers, AlGaAs claddings, and outer mirrors consisting of Ag. For the results shown in Fig. \ref{fig:loads}(b), there is also a 50 nm thick vacuum nanogap in the middle of the structure. The centre points of the emitter and absorber layers are located halfway between the mirrors closest to them and the center point of the full cavity (also other configurations have been experimented with, but this choice has provided the largest overall emission so far). (b) Reference structure, with a GaAs emitter layer emitting light into semi-infinite AlGaAs claddings.}
\label{fig:kaviteetit}
\end{figure}

To recap, the objective of this Subsection is to study whether the total emission of extremely thin layers can be boosted with cavity effects and by how much. To do that, Eq. (\ref{eq:Rrad}) is solved for structures illustrated in Fig. \ref{fig:kaviteetit}, with the structure in Fig. \ref{fig:kaviteetit}(a) representing a thin-film intracavity device including both an emitter and an absorber layer, and Fig. \ref{fig:kaviteetit}(b) representing a reference case, where a thin GaAs layer simply emits radiation into free space consisting of AlGaAs. In this Section, we always have a constant quasi-Fermi level separation of 1.3 eV and temperature of 300 K in the emitter layer. Similar to Ref. \citenum{kivisaari_2022}, the absorber layer in Fig. \ref{fig:kaviteetit}(a) is assumed to have significantly weaker excitation (lower applied bias and/or temperature), such that its own emission can be considered negligible. The permittivities deployed in the calculations are given in Refs. \citenum{palik_1998,glembocki_1998,palik_1998_add}, and the permittivity of Ag is multiplied by 100 similarly as in Ref. \citenum{kivisaari_2022} to decrease mirror losses and enable focusing on the optical processes that take place inside the cavity. Photon energies between 1.41 and 1.65 eV are considered, and the $K$ values span from 0 to 1.05 $N_{GaAs}k_0$, where $N_{GaAs}$ is the refractive index of GaAs and $k_0$ is the wave number in vacuum, both dependent on $\hbar\omega$.

\begin{figure}[htb]
\centering
\includegraphics[scale=0.7]{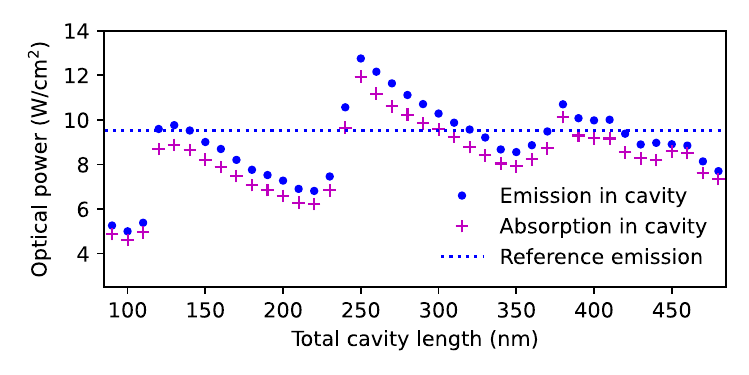}
\caption{$R_{rad}$ integrated over the emitter and absorber layers in the structure of Fig. \ref{fig:kaviteetit}(a) with emitter and absorber thicknesses of 20 nm and 60 nm (respectively), calculated at different total cavity lengths (datapoints for absorption technically show the integral of $-R_{rad}$ here). The figure also shows $R_{rad}$ integrated over the emitter layer in the reference structure of Fig. \ref{fig:kaviteetit}(b).}
\label{fig:em20}
\end{figure}

The dependence of the optical power on the cavity length is first illustrated in Fig. \ref{fig:em20} for the intracavity structure of Fig. \ref{fig:kaviteetit}(a) with a single example. Here, the thicknesses of the GaAs emitter and absorber layers are 20 and 60 nm, respectively. As in Ref. \citenum{kivisaari_2022}, the emitted and absorbed optical powers depend on the cavity length. Here, the maximum emission and absorption are reached at a total cavity length of approximately 250 nm. The optical power emitted by a similarly thick GaAs layer in the reference structure of Fig. \ref{fig:kaviteetit}(b) is also plotted as a constant dotted line in Fig. \ref{fig:em20}. It can be seen that at the maximum, the emission in the cavity is roughly 35 \% stronger than in the reference structure.

\begin{figure}[htb]
\centering
\includegraphics[scale=0.7]{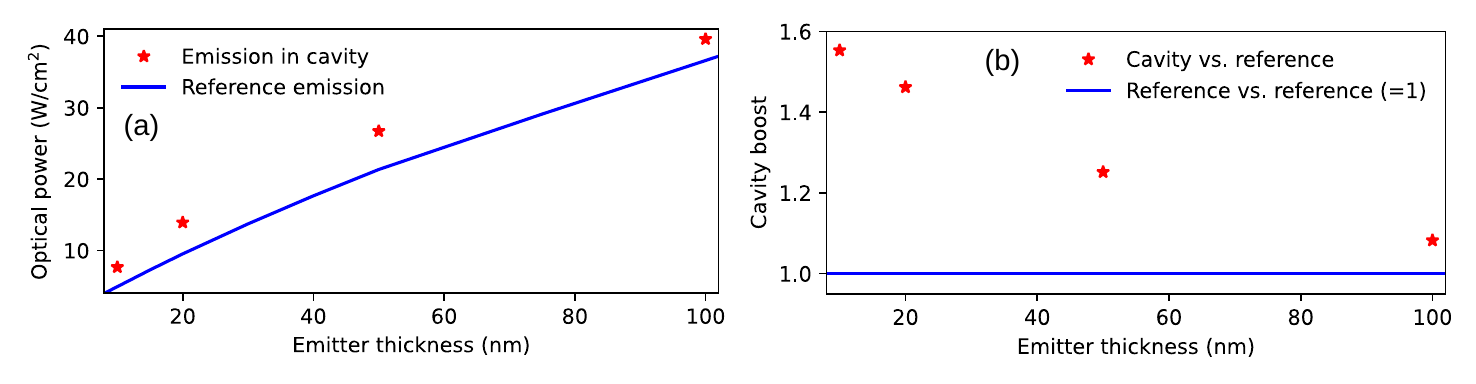}
\caption{(a) $R_{rad}$ integrated over the emitter layer in the intracavity structure of Fig. \ref{fig:kaviteetit}(a) and in the reference structure of Fig. \ref{fig:kaviteetit}(b). Emission in the intracavity structure is calculated with the cavity length and absorber thickness that maximize the emission. (b) Ratio of the optical power values plotted in (a), showing that the intracavity structure results in a 10 \% emission enhancement even at an emitter thickness of 100 nm.}
\label{fig:boosts}
\end{figure}

The intracavity emission enhancement illustrated in Fig. \ref{fig:em20} is generalized in Fig. \ref{fig:boosts}, where \ref{fig:boosts}(a) shows the maximum optical power emitted by the the GaAs layer in the cavity and the optical power emitted by the reference structure. Figure \ref{fig:boosts}(b) in turn shows the ratio of the values shown in Fig. \ref{fig:boosts}(a). For Fig. \ref{fig:boosts}(a), the values for emission in the cavity have been obtained by carrying out several calculations corresponding to Fig. \ref{fig:em20} and taking the maximum value from them (varying both total cavity length and absorber layer thickness). It can be seen in Fig. \ref{fig:boosts}(b) that the intracavity structure amplifies the emission of the thin GaAs layer as compared to the reference case especially at layer thicknesses below 50 nm. However, even for an emitter layer thickness of 100 nm, the boost from the cavity in Fig. \ref{fig:boosts}(b) is still roughly 10 \%. For more insight, Appendix \ref{sec:appendixc} presents spectral plots of the emission patterns in selected intracavity and reference structures.

\begin{figure}[htb]
\centering
\includegraphics[width=\textwidth]{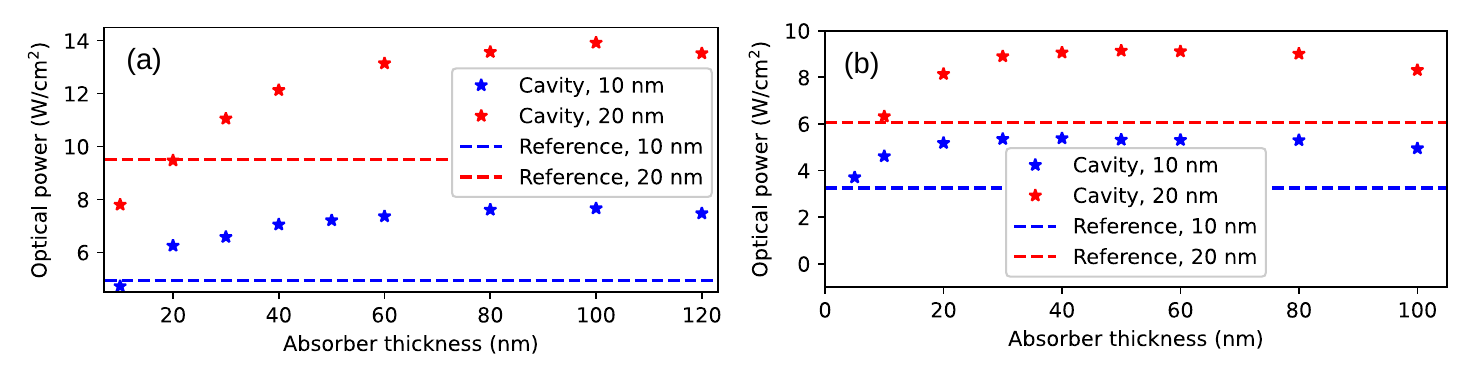}
\caption{(a) $R_{rad}$ integrated over the emitter layer in the intracavity structure as a function of the absorber layer thickness, with emitter layer thicknesses of 10 and 20 nm. All the values are shown for the cavity length that maximizes the emission, and $R_{rad}$ integrated for the reference structure is also shown. (b) Information from (a) calculated for the case where a 50 nm thick vacuum nanogap is inserted in the middle of the intracavity structure. Calculation of the reference values in (b) is explained in the main text.}
\label{fig:loads}
\end{figure}

To find the optimum values shown in Fig. \ref{fig:boosts}, both the cavity length and the absorber layer thickness were varied for each emitter layer thickness. Interestingly, to reach the maximum values shown in Fig. \ref{fig:boosts}, the optimum absorber layer thickness has to be somewhat larger than the emitter layer thickness. To illustrate this load matching aspect, Fig. \ref{fig:loads}(a) shows the maximum emission for 10 and 20 nm thick emitter layers in the cavity as a function of the absorber layer thickness for the structure illustrated in Fig. \ref{fig:kaviteetit}(a). Figure \ref{fig:loads}(b) shows the result for a similar structure but with a 50 nm thick vacuum gap inserted in the middle of the structure (necessary for thermophotonic applications). In both Figs. \ref{fig:loads}(a)-(b), the optimum absorber layer thickness is a few times the thickness of the emitter. Reference values for the vacuum gap case in Fig. \ref{fig:loads}(b) are calculated by taking the structure of Fig. \ref{fig:kaviteetit}(a), inserting the vacuum gap in the middle, removing the mirror on the left side, and using large enough thicknesses of the passive AlGaAs layers such that varying them does not change the integrated $R_{rad}$ any more. There, the left-side mirror was removed to cancel the resonant effect from photons reflecting from it. Other possible reference structures could have been the structure of Fig. \ref{fig:kaviteetit}(a) simply with a long cavity, or the reference structure of Fig. \ref{fig:kaviteetit}(b) with vacuum gaps inserted in both semi-infinite AlGaAs layers at relatively large distances from the GaAs layer.

\subsection{Effects of resonances \& scattering in ultra-thin-film solar cells} \label{sec:etsolarcells}

\begin{figure}[htb]
\centering
\includegraphics[width=0.28\textwidth]{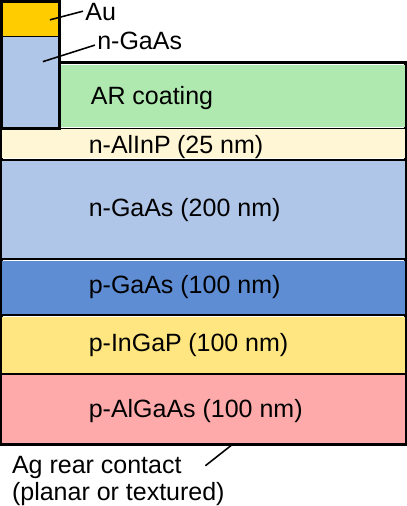}
\caption{Ultra-thin solar cell structure essentially similar to the one in Refs. \citenum{vaneerden_2019,vaneerden_2020}, to be simulated with the RTDD and IRTDD models in this Subsection. Later, 'specular' refers to simulations where the rear contact is assumed planar, and 'diffusively scattering' refers to simulations where it is assumed textured.}
\label{fig:solcell}
\end{figure}

To recap, the goal of this Subsection is to find out how much resonances and surface texturing affect the emission of moderately thin layers in thin-film devices interacting with free space. To do this, the ultra-thin solar cell illustrated in Fig. \ref{fig:solcell} is simulated as a model example with comparative RTDD and IRTDD simulations (note that even if the primary purpose of solar cells is to absorb light, studying their emissivity is one important part of optimizing their efficiency). The objective is therefore to study the role of resonant effects in the overall emission, first in the case where all the surfaces are perfectly planar. Lastly, another RTDD simulation is carried out where the rear surface is assumed to have an ideal texturation, scattering all incoming light homogeneously into all the modes propagating in the adjacent layer.

The simulation parameters are taken from Refs. \citenum{palik_1998,glembocki_1998,palik_1998_add,vurgaftman_2001,levinshtein_1996,levinshtein_1996_2,chen_1991,schubert_1995,linnik_2002,ozaki_1993,rodriguez_2017}, and solar light [resulting in $G_s$ in Eq. (\ref{eq:DDyhtalot})] enters the structure in normal incidence similarly as in Ref. \citenum{kivisaari_2021}, with the ASTM G173 solar intensity considered for photon energies between 1.35 and 4.43 eV. To calculate $R_{rad}$, photon energies between 1.35 and 1.65 eV are considered, and $K$ values span from 0 to $N_{GaAs}k_0$. Contribution from purely evanescent waves ($K>N_{GaAs}k_0$) is therefore not considered, as it is evident that they cannot be considered with the conventional RT model. The conventional RT simulation is carried out in the full semiconductor layer stack up to $K=N_{AlGaAs}k_0$, and only in the GaAs layers for $K$ values between $N_{AlGaAs}k_0$ and $N_{GaAs}k_0$, as they do not represent modes propagating in AlGaAs ($N_{AlGaAs}$ is the refractive index of AlGaAs). Outer boundary reflectances in the RTDD model are calculated with an equivalent of the transfer matrix method \cite{kivisaari_2018}.

\begin{figure}[htb]
\centering
\includegraphics[width=\textwidth]{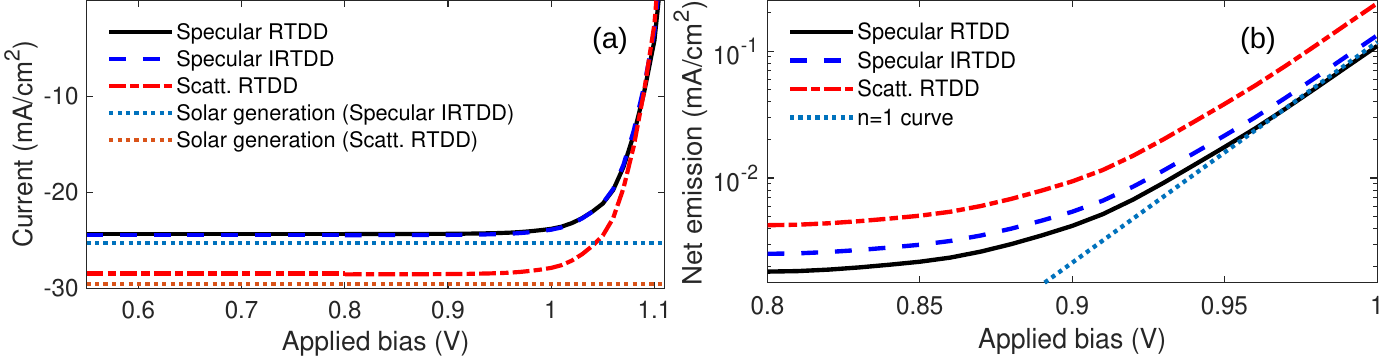}
\caption{(a) Current-voltage (IV) characteristics of the ultra-thin solar cell calculated with the specular RTDD, specular IRTDD, and diffusively scattering RTDD ('Scatt. RTDD') model. The solar generation rates (calculated by integrating $G_s$ of Eq. (\ref{eq:DDyhtalot}) over the whole device) are also shown in the figure ($G_s$ calculated with the specular IRTDD model is used also in the specular RTDD model to enable focusing only on the differences in $R_{rad}$). (b) Net emission rates under illumination calculated by integrating $R_{rad}$ over the whole device. The $J_{01}\exp(qV_a/k_BT)$ curve calculated from the applied bias is also shown in the figure (''$n=1$ curve'').}
\label{fig:iv}
\end{figure}

Beginning with the full-device characteristics under illumination, the simulated current-voltage (IV) curves are shown in Fig. \ref{fig:iv}(a) as calculated with the RTDD and IRTDD simulation. First of all, it can be seen that the IV curves from the specular RTDD and IRTDD simulations are practically equal with each other with the linear scale used in the figure. Therefore we conclude that the additional resonant effects included in the IRTDD simulation (illustrated more in Appendix \ref{sec:appendixd}) do not show up in the integrated device-level characteristics when plotted in the linear scale using the zoom level of Fig. \ref{fig:iv}(a). Also shown is the IV curve for the diffusively scattering RTDD case, where we observe that the short-circuit current density is 16 \% larger than for the specular case. The total solar generation rates calculated for the devices are also shown in the figure. The difference between the solar generation rates and short-circuit currents is caused by leakage of electrons generated in the AlGaAs layer in Fig. \ref{fig:solcell}. Later we find that running the simulations without the AlGaAs layer also removes the difference between short-circuit currents and solar generation rates observed here.

In Fig. \ref{fig:iv}(a), solar generation rate was calculated from IRT in both specular RTDD and IRTDD simulations to enable focusing strictly on differences in their emission. Calculating the solar generation rate instead from the RT model results in a difference of roughly 1 \%, as resonant effects are not accounted for (note that solar light is projected fully into the normally incident mode, and therefore in specular calculations there is no smearing out of resonances due to angle integration).

To focus more on the emission predicted by the different simulations, Fig. \ref{fig:iv}(b) shows $R_{rad}$ integrated over all the semiconductor layers (expressed in units of current density) under illumination, on a logarithmic scale. There, first of all we see that the net emission calculated in such a way is essentially equal between the RTDD and IRTDD simulations at the maximum power point of 1.0 V (the difference is roughly 10 \% in this example). At 0.8 V, on the other hand, the net emission rate is somewhat more different between the RTDD and IRTDD simulations. However, at 0.8 V, this net emission is only a tiny fraction of the total current of the solar cell under illumination and therefore essentially does not affect the total current. In the diffusively scattering RTDD case also shown in Fig. \ref{fig:iv}(b), the net emission is larger than in the calculations assuming specular reflections at all biases, as part of the photons emitted originally into large angles are able to scatter into the escape cone of air when reflecting from the textured rear contact.

\begin{figure}[htb]
\centering
\includegraphics[width=\textwidth]{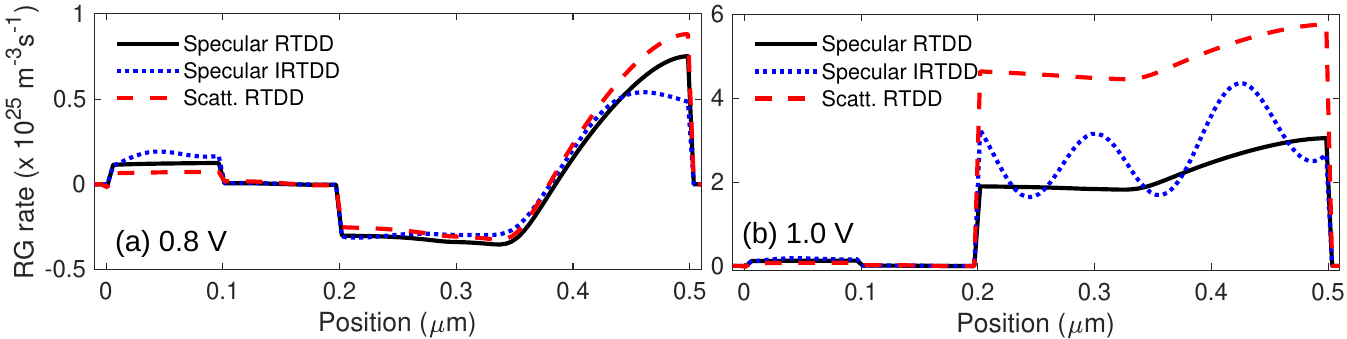}
\caption{$R_{rad}$ of Eq. (\ref{eq:DDyhtalot}) as a function of position at (a) 0.8 V and (b) 1.0 V, calculated under illumination with the specular RTDD, specular IRTDD, and diffusively scattering RTDD ('Scatt. RTDD') model. Absorption in the metal mirror to the left from AlGaAs is included in the calculations but not in $R_{rad}$ plotted here, as it does not introduce electron-hole pairs to the semiconductor layers.}
\label{fig:netrg}
\end{figure}

The net RG rates [$R_{rad}$ in Eq. (\ref{eq:Rrad})] discussed in the previous paragraph for applied biases 0.8 V and 1.0 V are illustrated in more detail in Fig. \ref{fig:netrg}. In Fig. \ref{fig:netrg}, $R_{rad}(z)$ is shown as a function of position under illumination at applied biases (a) 0.8 V and (b) 1.0 V. Note that the curves shown in Fig. \ref{fig:iv}(b) were obtained by integrating the curves of Fig. \ref{fig:netrg} over position. In Fig. \ref{fig:netrg}(a), $R_{rad}$ is positive close to the top surface of the solar cell and negative further away from it (towards the left in the figure). This is essentially the same net photon recycling process that we reported in Ref. \citenum{kivisaari_2021}. Interestingly, there is also non-negligible emission from the AlGaAs layer at the left side of the figure. This is due to sunlight absorbed by the AlGaAs layer, which results in a quasi-Fermi level separation of 1.23 eV there (see band diagrams in Appendix \ref{sec:appendixd}). This is associated with emission from AlGaAs even with the drift-diffusion currents removing excess electrons and holes towards lower-bandgap layers and contacts. 

Figure \ref{fig:netrg}(b) repeats the information of Fig. \ref{fig:netrg}(a) at the maximum power point of 1.0 V. There, it can be seen that the rates are roughly of the same magnitude in the specular IRTDD and RTDD simulation, apart from the interference pattern remaining in the IRTDD simulation. In the scattering RTDD case also plotted in Fig. \ref{fig:netrg}(b), $R_{rad}$ is larger due to the textured back mirror which allows also photons initially emitted to larger angles to escape the solar cell, as discussed above. Appendix \ref{sec:appendixd} gives more detailed information on the direction and energy dependence of the recombination-generation rates shown here.

\begin{figure}[htb]
\centering
\includegraphics[width=\textwidth]{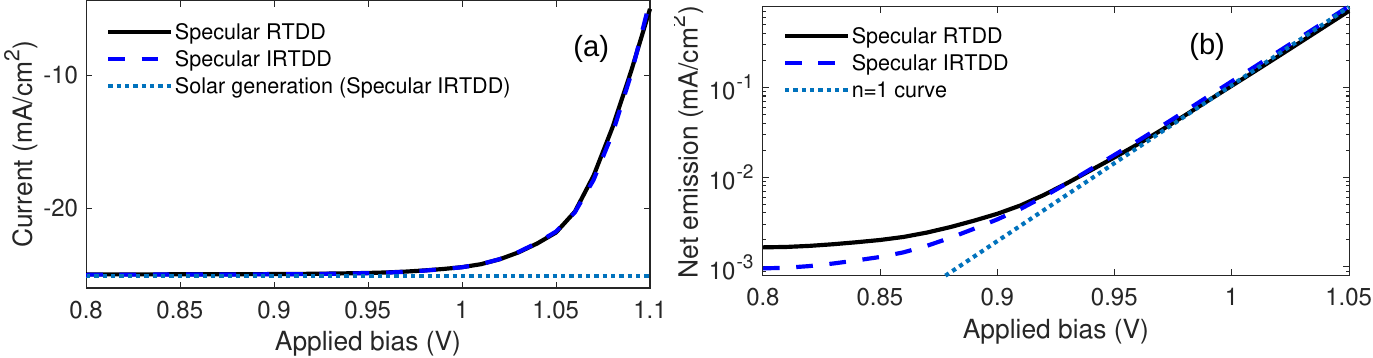}
\caption{Information of Fig. \ref{fig:iv} repeated for a case where the AlGaAs layer is removed (see the device structure in Fig. \ref{fig:solcell}). Scattering RTDD simulation has not been carried out for this device structure.}
\label{fig:iv_noalgaas}
\end{figure}

To check whether modifying the layer structure somewhat would cause differences between the RTDD and IRTDD simulations, Fig. \ref{fig:iv_noalgaas} repeats the information of Fig. \ref{fig:iv} for the device shown in Fig. \ref{fig:solcell}, but with the bottommost AlGaAs layer removed. First of all, it can be seen in Fig. \ref{fig:iv_noalgaas}(a) that the IV curves from the two different simulations are again essentially equal to each other (on linear scale and with this zoom level). Moreover, here the short-circuit current is practically equal to the solar generation rate expressed in mA/cm$^2$. Removing the AlGaAs layer therefore removes the related electron leakage in the simulations, even if a few percent of the incoming solar light is still absorbed in the GaInP layer next to the rear contact. The net emission rate is also shown for this case in Fig. \ref{fig:iv_noalgaas}(b), calculated similarly as in Fig. \ref{fig:iv}(b). Interestingly, also here there is a notable difference between the RTDD and IRTDD case only at the small voltages but not at the maximum power point of 1.0 V.

As a final note, if the active layer of the device is very close to an optical element that supports evanescent coupling (e.g. a metal mirror), and if correspondingly the higher-$K$ modes are considered in the IRTDD model, larger differences between the IRTDD and RTDD model are expected. But this is evident, as the conventional RT model is not suitable for studying waveforms that are completely evanescent.

\section{CONCLUSIONS}

In this work, we reported on our theoretical investigations of the role of resonances in two device types becoming more feasible thanks to recent developments in thin-film processing: extremely thin active layers placed in a cavity, and moderately thin layers in a thin-film device interacting with free space. Based on the results, the total emission of extremely thin layers can be boosted with resonant effects. However, it requires low-loss mirrors and, in the case of thermophotonic applications, high-quality vacuum gaps, all combined with a rather thin total device thickness. On the other hand, based on the results presented here, the total emission of moderately thin layers can be calculated very accurately even without considering wave-optical effects, as long as the reflectances of the outer boundaries are known as a function of energy and angle, and the active layer is not very close to optical elements supporting direct evanescent coupling. In addition, the results demonstrated how a corresponding radiative transfer approach coupled with the drift-diffusion model enables studying even diffusively scattering surfaces and their effect on the device performance. The results of this paper in part support efforts to develop simpler modeling tools for emerging thin-film devices. There, the aim would be to account for all the important effects to a sufficient quantitative accuracy while being able to directly study the effects of various design parameters on the overall device performance.

\acknowledgments

We acknowledge financial support from the European Union's Horizon 2020 programme (Grant Agreement Nos. 951976 and 964698). The calculations were performed using the computational resources provided by the Aalto Science-IT project.

\appendix

\section{Complementary definitions for the optical models} \label{sec:appendixa}

Here, the unified formulation of the RT and IRT models of Eq. (\ref{eq:RT_gen}) is completed in accordance with their more established forms. The RT model deployed in this work follows from casting the angle-dependent formulation of Ref. \citenum{kivisaari_2019} into being $K$-dependent as required by Eq. (\ref{eq:Rrad}). Beginning with the damping coefficient $\alpha(z,K,\omega)$, in RT it is the material's absorption coefficient $\alpha_0$ divided by an angular factor, i.e.,
\begin{equation}
\alpha(z,K,\omega) = \frac{\alpha_0(\omega)}{\cos\theta} = \frac{\alpha_0(\omega)}{\mathcal{K}},
\end{equation}
where $\theta$ is the propagation angle and $\mathcal{K} = \sqrt{1-\left[K/(Nk_0)\right]^2}$ ($N$ is the refractive index). The $\beta$ coefficient of Eq. (\ref{eq:RT_gen}) is simply equal to 0 in the RT model. In RT, $v$ is the speed of light in medium given by $v=c/N$, and the direction-dependent density of states is given by
\begin{equation}
\mathcal{G}_{RT}(z,K,\omega)=\frac{1}{(Nk_0)^2}g(z,\omega),
\label{eq:grt}
\end{equation}
where the prefactor is related to the change of integration variables from propagation angle $\theta$ (deployed in Ref. \citenum{kivisaari_2019}) into $K$, and $g(z,K,\omega)$ is the customary bulk 3-dimensional optical density of states given by
\begin{equation}
g(z,\omega) = \frac{(\hbar\omega)^2}{2\pi^2\hbar^3v^3}.
\end{equation}
With this, the net recombination-generation rate in Eq. (\ref{eq:Rrad}) matches the one reported in Ref. \citenum{kivisaari_2019}.

For IRT, the more complex damping and scattering coefficients $\alpha(z,K,\omega)$ and $\beta(z,K,\omega)$ ensure that the IRT equation provides a rigorous solution to Maxwell's equations within the layer structure. Both $\alpha$ and $\beta$, as well as the transmission matrices required at interfaces between different materials, are calculated from the dyadic Green's functions as detailed in Refs. \citenum{kivisaari_2021,partanen_2017_irt}. In this paper, the dyadic Green's functions required for $\alpha$ and $\beta$ etc. are calculated by using the optical admittance method introduced in Ref. \citenum{kivisaari_2018}. The direction-dependent density of states in the FED and therefore also in the IRT calculations is given by
\begin{equation}
\mathcal{G}_{FED/IRT}=\frac{2\pi}{\hbar}\rho(z,K,\omega),
\label{eq:girt}
\end{equation}
where $\rho(z,K,\omega)$ is the local density of states calculated from dyadic Green's functions and likewise specified in the supplementary material of Ref. \citenum{partanen_2017_irt} [note that $\rho(z,K,\omega)$ retains $K$-dependence and has different units from $g(z,\omega)$ -- hence also the rather different prefactors in Eqs. (\ref{eq:grt}) and (\ref{eq:girt})]. The generalized speed of light $v$ in IRT is given for TE and TM polarizations in Appendix A of Ref. \citenum{kivisaari_2021} (the special definition of $v$ in the IRT model is a mathematical means to handle its boundary conditions, as explained in Ref. \citenum{kivisaari_2021}; in the FED formulation of Section \ref{sec:cavities}, $v$ is still given by $c/N$). Substituting the above into Eq. (\ref{eq:Rrad}) leads to the same formulation of the net recombination-generation rate as in Ref. \citenum{kivisaari_2021}.

Quick note on the plots to be shown in Appendices \ref{sec:appendixc} and \ref{sec:appendixd}: direction-dependent plots of emission are usually easier to interpret as a function of the propagation angle $\theta$ than as a function of $K$. Therefore, for those plots, we make use of the following dependencies: $dK = Nk_0\cos\theta d\theta$ and $K = Nk_0\sin\theta$ (the factors denoting direction in Eq. (\ref{eq:Rrad})]. A meaningful quantity to be plotted as a function of angle is therefore given by $(Nk_0)^2\sin\theta\cos\theta r(z,\theta,\omega)$, where $r(z,\theta,\omega)$ is simply $r(z,K,\omega)$ calculated for the corresponding $\theta$. Phenomenologically, the $\cos\theta$ term here represents the projection of emission into the $z$ direction, and the $\sin\theta$ term accounts for the larger optical density of states available at larger angles. In the plots of Appendix \ref{sec:appendixc}, the $\sin\theta$ factor is left out to focus on the directivity of the emission without the effect from the optical density of states. The plots of Appendix \ref{sec:appendixd} include both $\cos\theta$ and $\sin\theta$ factors similarly as in Ref. \citenum{kivisaari_2021}. Additionally, as all the direction-dependent plots of Appendices \ref{sec:appendixc} and \ref{sec:appendixd} are normalized and shown in arbitrary units, the $(Nk_0)^2$ term is left out.

\section{Specular and diffusive scattering in RT} \label{sec:appendixb}

In addition to comparing specular IRTDD and RTDD models, Section \ref{sec:etsolarcells} also presents RTDD results obtained for textured surfaces where light scatters evenly in all directions (perfectly diffusive scattering). To formulate the specular and diffusive scattering cases, the total spectral radiance is constructed from the forward- and backward-propagating photon numbers $\phi^+(z,K,\omega)$ and $\phi^ -(z,K,\omega)$. In the special case of an ultrathin solar cell where we assume a rear contact at the left side of the domain at $z=0$ (Ag rear contact in Fig. \ref{fig:solcell}), we are interested in the value of $\phi^-$ there. Let us denote this as $\phi^-(0,K,\omega)\equiv\hat{\phi}^-(K,\omega)$. The reflected photon number is then given by 
\begin{equation}
\hat{\phi}^+(K,\omega)=\mathcal{R}\hat{\phi}^-(K,\omega),
\end{equation}
where $\mathcal{R}$ is a reflection operator. For specular reflection, it simply multiplies the photon number at each angle by a reflectivity between 0 and 1 to give a boundary value for $\phi_{\uparrow}$ for the corresponding angle. In other words, for specular reflection one gets
\begin{equation}
\mathcal{R}\hat{\phi}^-(K,\omega)=R(K,\omega)\hat{\phi}^-(K,\omega),
\label{eq:specularR}
\end{equation}
where $R(K,\omega)$ is the reflectance of the mirror.

For the perfectly diffusive scattering case, we assume that for each $K$, the leftward photon number is reflected equally into all directions propagating in the adjacent layer (with the corresponding maximum $K$ denoted as $K_{max}=N_{AlGaAs}k_0$). The corresponding boundary value $\hat{\phi}^+$ is constant for $K\in(0,K_{max})$, and it can be denoted as $\phi_{const}$. The value of $\phi_{const}$ can be derived by setting the total radiance corresponding to it equal to the total radiance hitting the rear contact multiplied by reflectivity of the contact:
\[
s_{const} = s_{refl}
\]
\begin{equation}
\int_0^{K_{max}}dK\mathcal{G}(z,K,\omega)\phi_{const}(\omega)= \int_0^{K_{max}}dKR(K,\omega)\mathcal{G}(z,K,\omega)\hat{\phi}^-(K,\omega).
\end{equation}
Here, a possible zeroth-order approximation for $R(K,\omega)$ is to use the one calculated for a planar interface. Solving $\phi_{const}$ from this, we get
\begin{equation}
\phi_{const}(\omega)=\frac{2}{K_{max}^2}\int_0^{K_{max}}dKR(K,\omega)\hat{\phi}^-(K,\omega)K.
\label{eq:phiconstrt}
\end{equation}
For simplicity, in the calculations of this paper, we use an angle-independent $R(\omega)$ in Eq. (\ref{eq:phiconstrt}) calculated for normal incidence from an equivalent of the scattering matrix method for a planar Ag/AlGaAs interface.

A similar perfectly diffusive reflection can be applied for the incoming solar light. Let us denote the incoming solar radiance hitting the rear contact by $P_0(\omega)$, where subscript 0 refers to $K=0$, as the incoming solar radiance is fully projected to the normal mode. To make this reflect equally to all directions at the rear contact, we seek an angle-independent boundary value for reflecting solar light $\phi_{solar}(\omega)$, which would integrate to the same radiance as $R(\omega)P_0(\omega)$, where $R(\omega)$ is the reflectance of the rear contact:
\begin{equation}
\frac{1}{2}\frac{(\hbar\omega)^3}{\pi^2\hbar^3v^2}\phi_{solar}(\omega)\frac{1}{n_r^2k_0^2}\int_0^{K_{max}}dK K = R(\omega)P_0(\omega).
\end{equation}
Substituting $v=c/N$ and solving for $\phi_{solar}$, this results in
\begin{equation}
	\phi_{solar}(\omega)=\frac{4\pi^2\hbar^3c^2}{(\hbar\omega)^3}\frac{k_0^2}{K_{max}^2}R(\omega)P_0(\omega).
	\label{eq:phisolar}
\end{equation}
Also here, for $R(\omega)$ we use the reflectance calculated for a planar Ag/AlGaAs interface at normal incidence from an equivalent of the scattering matrix method. The value for $\phi_{solar}$ given by Eq. (\ref{eq:phisolar}) is used as the input value in an additional direction-dependent RT equation (likewise with the diffusively scattering boundary condition) that tracks the propagation, absorption, and consecutive scattering of the solar light initially scattered from the rear contact. Such a separation of the total photon number in Eq. (\ref{eq:RT_gen}) into internally emitted and externally incident photon numbers $\phi_{internal}$ and $\phi_{external}$ (such that the total photon number is $\phi=\phi_{internal}+\phi_{external}$) can be done thanks to the additivity of the $d/dz$ operator. In such a way, only the RT equation for internally emitted photons includes the $\eta$ source term.


\clearpage

\section{Complementary results for intracavity structures} \label{sec:appendixc}

\begin{figure}[htb]
\centering
\includegraphics[scale=0.7]{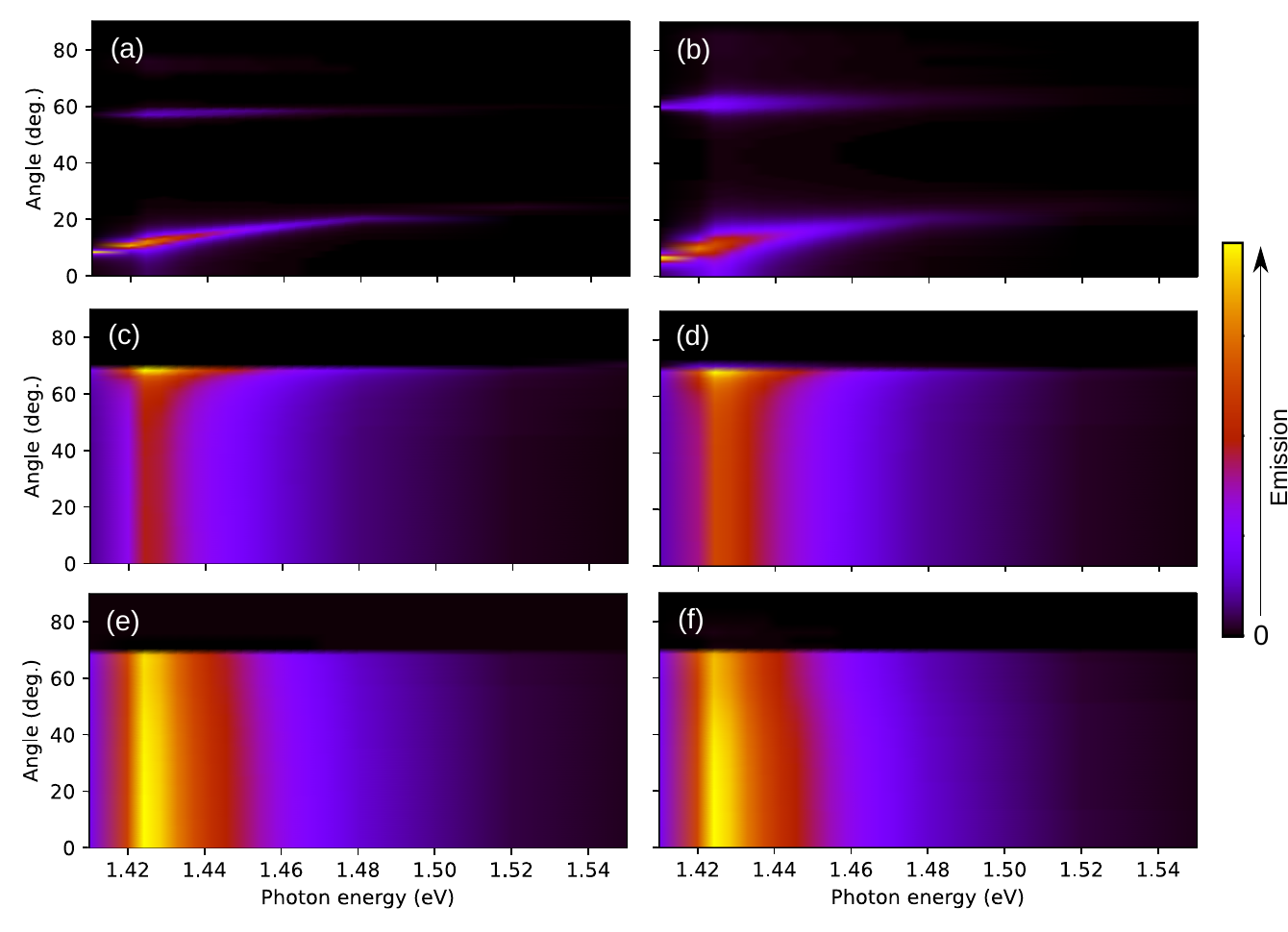}
\caption{Spectral RG rate $\cos\theta r(z,\theta,\omega)$ (a.u., see Appendix \ref{sec:appendixa}) averaged over the emitter layer for (a) intracavity structure of Fig. \ref{fig:kaviteetit}(a) with emitter 10 nm and absorber 50 nm, (b) intracavity structure of Fig. \ref{fig:kaviteetit}(a) with emitter 100 nm and absorber 120 nm, and (c)--(f) reference structure of Fig. \ref{fig:kaviteetit}(b) with emitter thicknesses 10, 50, 100, and 300 nm, respectively. The rates are plotted as a function of photon energy and propagation angle in GaAs.}
\label{fig:spectral}
\end{figure}

To complement the results presented in Section \ref{sec:cavities}, Fig. \ref{fig:spectral} shows spectral recombination rates $\cos\theta r(z,\theta,\omega)$ [see Eq. (\ref{eq:Rrad}) and last paragraph of Appendix \ref{sec:appendixa}] averaged over the emitter layer for different cases. First, Fig. \ref{fig:spectral}(a) corresponds to the intracavity structure of Fig. \ref{fig:kaviteetit}(a) with emitter and absorber layer thicknesses of 10 and 50 nm, and Fig. \ref{fig:spectral}(b) corresponds to the intracavity structure of Fig. \ref{fig:kaviteetit}(a) with emitter and absorber layer thicknesses of 100 and 120 nm. Then, Figs. \ref{fig:spectral}(c)--(f) correspond to the reference structure of Fig. \ref{fig:kaviteetit}(b) with emitter layer thicknesses of 10, 50, 100, and 300 nm, respectively. Comparing Figs. \ref{fig:spectral}(a)--(b), it can be seen that light is emitted into essentially discrete angles similarly as in Ref. \citenum{kivisaari_2022}, but with the emission peaks broadening as a function of angle as the emitter and absorber layer thicknesses are increased [Fig. \ref{fig:spectral}(b)]. For the reference structure plotted in Figs. \ref{fig:spectral}(c)--(f), light is emitted over a constant spectrum into all angles up to roughly 70 degrees, above which the modes do not propagate in AlGaAs due to its smaller refractive index. In Figs. \ref{fig:spectral}(c)--(d), we also see that if the emitter layer is extremely thin, the emission in the reference case has its peak at the largest angles due to minimal reabsorption of emitted photons even when the propagation angle is large.

\clearpage

\section{Complementary results for ultra-thin solar cells} \label{sec:appendixd}

\begin{figure}[htb]
\centering
\includegraphics[width=\textwidth]{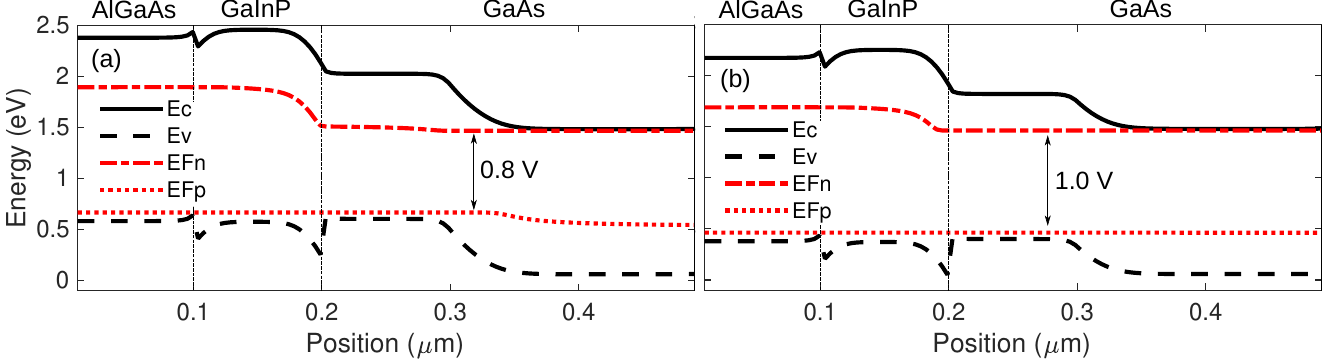}
\caption{Band diagram of the structure of Fig. \ref{fig:solcell} from the specular IRTDD simulation under illumination at (a) 0.8 V and (b) 1.0 V, including the AlGaAs, GaInP, and GaAs layers. The contact boundaries (where the quasi-Fermi levels would merge) are located outside the region shown in the plot.}
\label{fig:bd}
\end{figure}

\begin{figure}[htb]
\centering
\includegraphics[width=\textwidth]{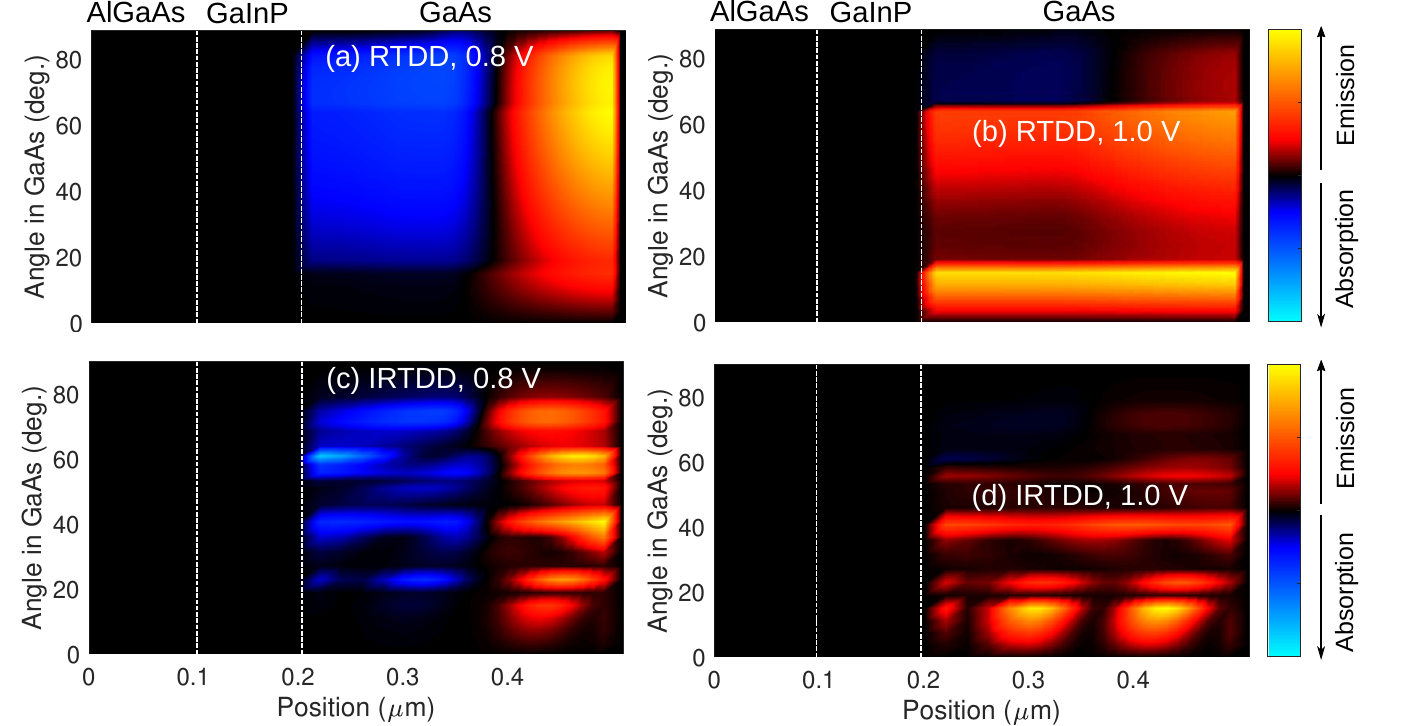}
\caption{Ultra-thin solar cell simulations with specular reflections under solar illumination: Spectral RG rate $\sin\theta\cos\theta r(z,\theta,\omega)$ (a.u., see last paragraph of Appendix \ref{sec:appendixa}) as a function of position and angle for (a) RTDD model at 0.8 V, (b) RTDD model at 1.0 V, (c) IRTDD model at 0.8 V, and (d) IRTDD model at 1.0 V. The photon energy is 1.44 eV. Each subfigure has been normalized by its own maximum value.}
\label{fig:rgangle}
\end{figure}

The band diagrams of the AlGaAs, GaInP, and GaAs layers of Fig. \ref{fig:solcell} are shown in Fig. \ref{fig:bd}, with applied biases of (a) 0.8 V and (b) 1.0 V. The band diagrams are from the specular IRTDD calculation, and the contact boundaries (where the quasi-Fermi levels would merge) are not shown in the plot. In both Figs. \ref{fig:bd}(a) and (b), the quasi-Fermi level separation is larger than the applied bias (marked inside the figures) in selected parts of the device thanks to the position-dependent solar generation rate $G_s$. In particular, due to the small thickness of the solar cell, notable part (roughly 7 \%) of the incoming solar light is absorbed by the AlGaAs and GaInP layers, and this elevates the quasi-Fermi level separation there to roughly 1.23 eV.

The spectral RG rate in the ultra-thin solar cell of Fig. \ref{fig:solcell} is illustrated in Fig. \ref{fig:rgangle}, which shows $\sin\theta\cos\theta r(z,\theta,\omega)$ [see Eq. (\ref{eq:Rrad}) and last paragraph of Appendix \ref{sec:appendixa}] at a photon energy of 1.44 eV as a function of angle under illumination from (a) RTDD simulation at 0.8 V, (b) RTDD simulation at 1.0 V, (c) IRTDD simulation at 0.8 V, and (d) IRTDD simulation at 1.0 V (all simulated with specular reflections). The 0.8 V applied bias case illustrated in Fig. \ref{fig:rgangle}(a) and (c) represents a case where the current is essentially equal to the short-circuit current. There, the solar cell exhibits a similar net photon recycling process as in Ref. \citenum{kivisaari_2021}. The IRTDD case shown in Fig. \ref{fig:rgangle}(c) furthermore clearly exhibits resonant effects due to reflections from the rear mirror and the top surface.

The 1.0 V applied bias case shown in Figs. \ref{fig:rgangle}(b) and (d) represents the maximum power point of the solar cell. There, the largest spectral RG rate can be found below roughly 16 degrees, and it corresponds to light emitted into air. The RG rate between 16 degrees and 65 degrees corresponds to light emitted towards the rear contact where it is absorbed, and the RG rate above 65 degrees is the remaining net photon recycling inside the active GaAs layer. Also in Fig. \ref{fig:rgangle}(d), the IRTDD simulation exhibits resonant effects.

\begin{figure}[htb]
\centering
\includegraphics[width=\textwidth]{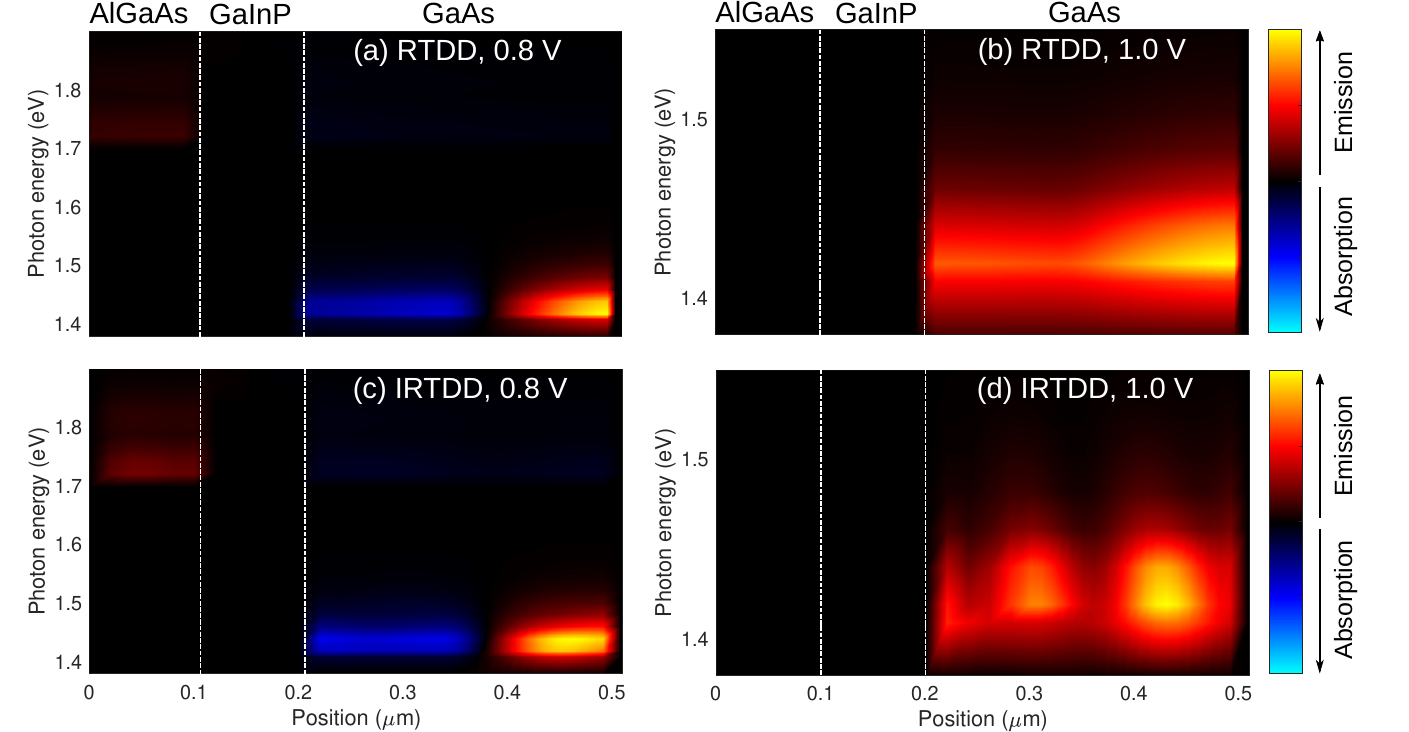}
\caption{Ultra-thin solar cell simulations with only specular reflections: RG rate (a.u.) as a function of photon energy and position for (a) RTDD model at 0.8 V, (b) RTDD model at 1.0 V, (c) IRTDD model at 0.8 V, and (d) IRTDD model at 1.0 V, with the angle (or $K$) integration already carried out. In (a) and (c), one can see faint emission also from the AlGaAs layer at photon energies above 1.7 eV due to the absorption of solar light in AlGaAs that creates a non-negligible quasi-Fermi level separation there [see band diagram in Fig. \ref{fig:bd}(a)]. In (b) and (d), emission in GaAs is powerful enough such that emission is AlGaAs would not be visible when normalized to the emission from GaAs. Therefore, the photon energy axis has been narrowed down in (b) and (d). Each subfigure has been normalized by its own maximum value.}
\label{fig:rgenergy}
\end{figure}

In Fig. \ref{fig:rgenergy}, the RG rates are shown for the cases corresponding to Fig. \ref{fig:rgangle}, but instead as a function of photon energy, such that the angle (or $K$) integration is already carried out. Also here, the RTDD and IRTDD simulations for the 0.8 V case in Figs. \ref{fig:rgenergy}(a) and (c) show the net photon recycling process taking place in the GaAs layer. Interestingly, they also show faint emission in the AlGaAs layer due to the quasi-Fermi level separation of 1.23 eV created by the absorption of solar light there (see band diagram in Fig. \ref{fig:bd}). The RG rate as a function of position and photon energy is shown at the maximum power point (applied bias 1.0 V) in Figs. \ref{fig:rgenergy}(b) and (d). At the maximum power point, there is notable emission only from GaAs, and the y axis has been zoomed accordingly. Also when plotted as a function of energy, the IRTDD simulation at the maximum power point in Fig. \ref{fig:rgenergy}(d) exhibits resonance effects.

\begin{figure}[htb]
\centering
\includegraphics[width=\textwidth]{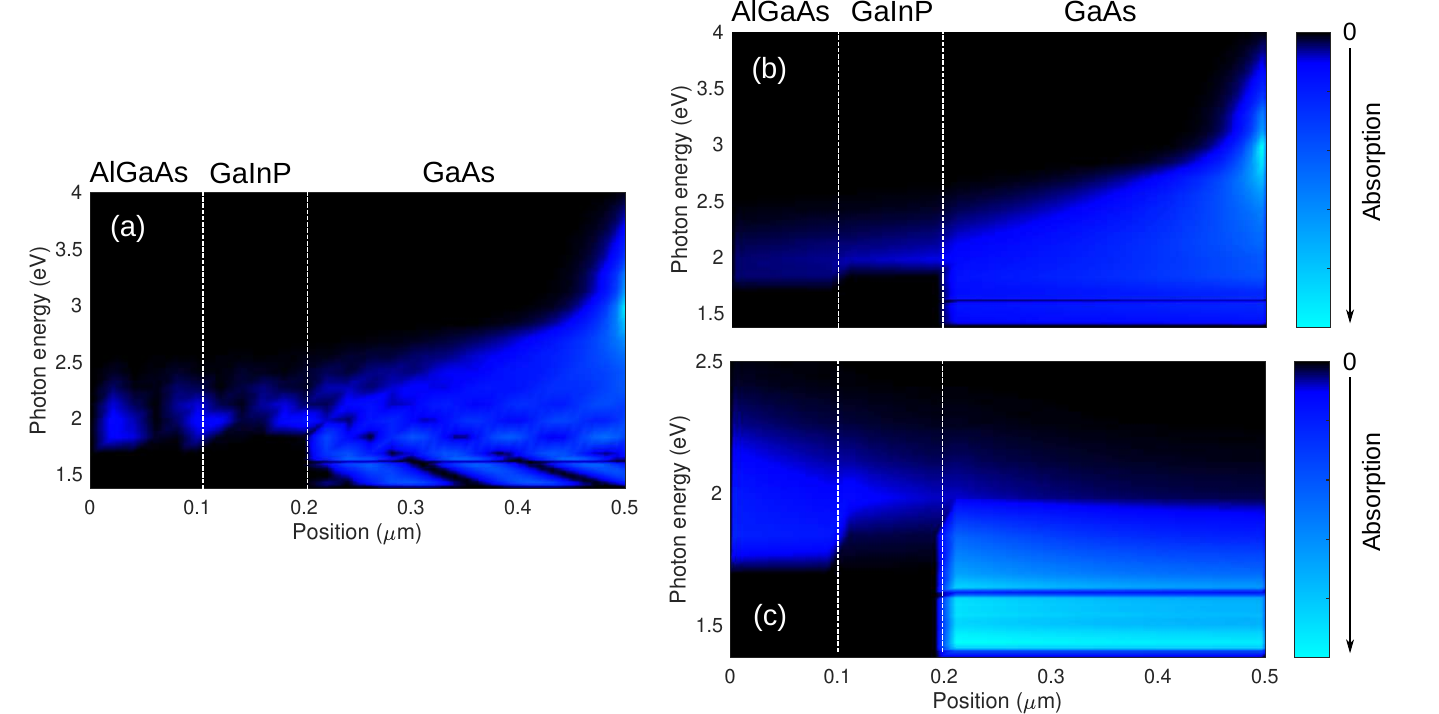}
\caption{Ultra-thin solar cell simulations: solar generation rate (a.u.; negative values denote absorption in analogy with other figures, i.e., integrating these over energy would result in $-G_s$) as a function of position and energy (a) in the specular IRT case, (b) in the scattering RT case for directly incident sunlight, and (c) in the scattering RT case for other directions than the normal incidence. Each subfigure has been normalized by its own maximum value.}
\label{fig:solabs}
\end{figure}

The absorption rate of solar light as a function of position and photon energy is illustrated for the specular and diffusive cases in Fig. \ref{fig:solabs}. There, Fig. \ref{fig:solabs}(a) shows the solar generation rate calculated with the specular IRT model. The $G_s$ calculated from this is deployed also in the specular RTDD simulation to enable focusing strictly on the differences in their emission with an equal excitation (difference between the solar generation rates calculated from specular IRT and RT simulations is roughly 1 \%, as mentioned previously). The solar generation rate in Fig. \ref{fig:solabs}(a) shows a clear resonance pattern at photon energies below roughly 2.5 eV, and there is considerable absorption of solar light also in the AlGaAs and GaInP layers. The solar generation rate from the scattering RT simulation is illustrated in Figs. \ref{fig:solabs}(b) and (c), where Fig. \ref{fig:solabs}(b) shows the single-pass generation rate of incoming photons in the incident 0 degree angle, and Fig. \ref{fig:solabs}(c) shows the generation rate due to solar photons scattered into other angles at the textured rear contact. In Figs. \ref{fig:solabs}(b)--(c), it can be seen that only the photons at energies below roughly 2.5 eV make it to the rear contact from which they can scatter to all angles (the y axis has been zoomed accordingly).

\begin{figure}[htb]
\centering
\includegraphics[width=\textwidth]{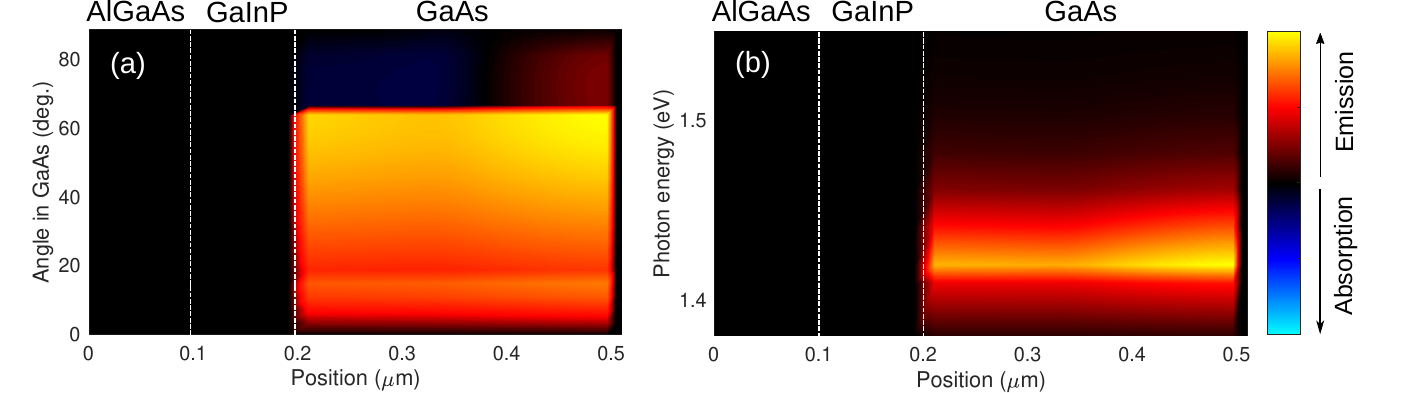}
\caption{Ultra-thin solar cell simulations with the diffusively scattering RTDD model: (a) Spectral RG rate $\sin\theta\cos\theta r(z,\theta,\omega)$ (a.u., see last paragraph of Appendix \ref{sec:appendixa}) as a function of position and angle with $\hbar\omega=1.44$ eV, and (b) RG rate (a.u.) as a function of position and energy from the scattering RT simulation at 1.0 V, after carrying out the angle (or $K$) integration. Both subfigures have been normalized by their own maximum values.}
\label{fig:rgscatt}
\end{figure}

The recombination-generation rate $\sin\theta\cos\theta r(z,\theta,\omega)$ [see Eq. (\ref{eq:Rrad}) and last paragraph of Appendix \ref{sec:appendixa}] from the diffusively scattering RT simulation is shown in Fig. \ref{fig:rgscatt}(a) at an applied bias of 1.0 V as a function of angle and position at photon energy 1.44 eV. The plot in Fig. \ref{fig:rgscatt}(a) can be compared with the specular RTDD simulation shown in Fig. \ref{fig:rgangle}(b). With that comparison, the emission into angles above roughly 40 degrees is accentuated in Fig. \ref{fig:rgscatt}(a), as part of those photons are scattered at the rear contact into angles smaller than $\sim$16 degrees where they can escape the solar cell. This is also behind the larger net emission rate in the scattering RT case observed in Figs. \ref{fig:iv}(b) and \ref{fig:netrg}(b). The angle (or $K$) integrated net recombination-generation rate as a function of photon energy in Fig. \ref{fig:rgscatt}(b) is not very different from the corresponding plot for the specular RTDD case shown in Fig. \ref{fig:rgenergy}(b), other than that the rate in Fig. \ref{fig:rgscatt}(b) is somewhat more evenly spread over the whole GaAs layer.


\end{document}